\documentstyle[12pt]{article}

\newcommand{\be}{\begin{equation}}
\newcommand{\ee}{\end{equation}}
\newcommand{\dd}{\partial}
\newcommand{\bea}{\begin{eqnarray}}
\newcommand{\eea}{\end{eqnarray}}

\begin{document}
\baselineskip .25in
\newcommand{\numero}{hep-th/9608023, SHEP 96/18}  

\newcommand{\titre}{Anti--Field Formalism and Non--Abelian Duality}
\newcommand{\auteura}{P. J. Hodges } 
\newcommand{\auteurb}{Noureddine Mohammedi}

\newcommand{\placea}{Department of Physics,\\University of
Southampton,\\Southampton, SO17 1BJ, \\ U.K. }
\newcommand{\placeb}{Universit\'{e} de Tours, Parc de Grandmont,F-37200
 Tours,France.}
\newcommand{\beq}{\begin{equation}}
\newcommand{\eeq}{\end{equation}}

\newcommand{\abstrait}
{The act of implementing non-Abelian duality in two dimensional sigma
models results unavoidably in an additional reducible symmetry.  The
Batalin--Vilkovisky formalism is employed to handle this new
symmetry. Valuable lessons are learnt here with respect to non--Abelian
duality.  We emphasise, in particular, the effects of the ghost sector
corresponding to this symmetry on non--Abelian duality.}
\begin{titlepage}
\hfill \numero  \\
\vspace{.5in}
\begin{center}
{\large{\bf \titre }}
\bigskip \\ \auteura 
\,\,\footnote{e-mail: pjh@hep.phys.soton.ac.uk}
 
    \bigskip  and
\bigskip \\
\auteurb  \footnote{e-mail:
nouri@hep.phys.soton.ac.uk.}\,\footnote{ Permanent address after
September 1996: \placeb}
\bigskip \\ 
\placea  \bigskip \\
\vspace{.9 in} 
{\bf Abstract}
\end{center}
\abstrait
 \bigskip \\
\end{titlepage}
\newpage
\section{Introduction}
Duality transformations have understandably brought about a surge of new
interests in string theory.  The importance of these transformations
lies in their ability to connect seemingly different string backgrounds.
This might shed some light on one of the longstanding problems in
superstring theory, namely the non--uniqueness of the low energy physics
expected from this theory.  As it is well--known, the phenomenology
predicted by superstring theory depends upon the way the extra six
dimensions are compactified.  Hence, if the spaces on which one carries
out the compactification are related to each other by duality
transformations, then their corresponding low energy physics should also
be related.  This is also the idea behind mirror symmetry \cite{yau92}
which might well be another manifestation of duality transformations
\cite{strom96,yin96}.
\par
The duality transformation that concerns us here is the so--called
T--duality \cite{busch88}.  This can be understood as canonical
transformations on the phase space of a sigma model \cite{lozan95}.
There is, however, a well defined procedure at the level of the
Lagrangian which allows the construction of dual theories
\cite{rocek92}.  It consists in gauging an isometry group of a
non--linear sigma model and at the same time restricting, by means of a
Lagrange multiplier, the gauge field to be pure gauge.  The integration
over the gauge fields (without a kinetic term) leads to the dual theory.
\par
The duality transformation is termed Abelian or non--Abelian depending
on whether the isometry group is Abelian or not.  Abelian duality has
proved to be of crucial importance in string \cite{giveon94} and
membrane \cite{schw96} theories.  On the other hand, its non--Abelian
counterpart has not yet been fully exploited \cite{frid84}.  This is
because non--Abelian duality is hampered by conceptual problems (such as
global issues and the fact that carrying out the transformation twice
does not lead to the original model).  One of the issues in non--Abelian
duality is the appearance, as explained below, of a new local symmetry
in the formlism \cite{moha96}.
\par
It is the aim of this paper to deal with the quantisation of this new
symmetry.  The understanding of this symmetry is crucial to any possible
exploitation ( and probably to the understanding of the other issues) of
non--Abelian duality.  We outline below the manifestation of this
symmetry.  As this symmetry is reducible we appeal to the
Batalin--Vilkovisky formalism \cite{bata85} for a rigourous treament.
The formalism is briefly summarised in section two.  Our main result is
that the dual theory depends on the ghost sector corresponding to this
new symmetry and on its gauge fixing conditions.  Let us therefore start
by stating the problem.
\par
Suppose that one has a two-dimensional theory described by an action
$S\left(\varphi\right)$ which is invariant under some global symmetry
for the generic fields $\varphi$.  Let us also assume that the
generators of this symmetry form a closed Lie algebra
${\mathcal{G}}$. Furthermore it is also assumed that one can gauge these
symmetry in an anomaly-free way. It is then straightforward to find the
dual of this theory at the classical level.  This is found by
considering the gauge invariant action \cite{rocek92}
\bea
I\left(\varphi,A, \Lambda\right)&=& S\left(\varphi,A\right)+\int
{\mathrm {d}}^2x {\mathrm {tr}}\left(\Lambda F
\right)\nonumber\\
F&\equiv& \epsilon^{\mu\nu}F_{\mu\nu}\,\,\,.
\label{firstaction}
\eea
Here  $S\left(\varphi,A\right)$ is the gauged version of 
$S\left(\varphi\right)$. The gauge field $A_\mu$ takes value in 
the Lie algebra ${\mathcal{G}}$ and 
$F_{\mu\nu}=\dd_\mu A_\nu -\dd_\nu A_\mu + 
\left[A_\mu\,\,,\,\, A_\nu\right]$ is the corresponding
field strength. The trace ${\mathrm{tr}}$ is 
the invariant bi-linear form of the 
Lie algebra ${\mathcal{G}}$ such that ${\mathrm{tr}}\left(XY\right)
=\eta_{ab}X^aY^b$.
\par
The new field $\Lambda$ is a Lagrange multiplier which, at the 
classical level, imposes the constraints
$F_{\mu\nu}=0$. This is then solved by $A_\mu = g^{-1}\dd_\mu g$,
where $g$ is an element in the Lie group corresponding to
${\mathcal{G}}$.  Recall now that $A_\mu$ and $\Lambda$ transform as 
\bea
A_\mu &\longrightarrow& h A_\mu h^{-1}
-\dd_\mu h h^{-1} \nonumber\\
\Lambda &\longrightarrow& h \Lambda h^{-1}
\label{usualsym}
\eea
where $h$ is the Lie algebra valued gauge function.  Of course, the
transformation of the generic field $\varphi$ is also governed by this
same function.  Using this gauge freedom, we can choose a gauge such
that $g=1$. Hence, in this gauge, the gauge field vanishes and the
action $I\left(\varphi,A, \Lambda\right)$ is classically equivalent to
the original action $S\left(\varphi\right)$.
\par
At the classical level, the dual theory is obtained by keeping the
Lagrange multiplier and eliminating instead the gauge fields by their
equations of motion.  We are supposing that the gauge fields appear
quadratically at most and without derivatives in the gauged action
$S\left(\varphi,A\right)$.  To get the right degrees of freedom in the
dual theory a gauge fixing condition must be chosen.
\par
The issues that concerns us in this paper are those necessary to
implement the duality transformation at the quantum level. This is a
well-known procedure if the Lie algebra ${\mathcal{G}}$ is
Abelian. However, if ${\mathcal{G}}$ is non-Abelian then the matter must
be considered carefully.  This is mainly because the action
(\ref{firstaction}) now has another local symmetry which must be taken
into account in the path integral. Due to the properties of the trace,
the gauge invariant action $I$ is also invariant under \cite{moha96}
\bea
\Lambda &\longrightarrow& \Lambda +
\left[\xi\,\,,\,\,F\right]
\nonumber\\
A_\mu &\longrightarrow& A_\mu\,\,\,\,,\,\,\,\,
\varphi \longrightarrow \varphi\,\,\,,
\label{newsym}
\eea
where $\xi$ is the new local gauge function corresponding to this extra
symmetry.  It should be noted that if the gauge function $\xi$ takes
value in the centre (or maximal ideal) of the Lie algebra
${\mathcal{G}}$, then the transformation of $\Lambda$ vanishes; thus the
new symmetry is reducible (i.e., not all the components of $\Lambda$
enter the transformation). This fact will have consequences, as we will
see, on the Faddeev-Popov ghosts required to gauge fix this new
symmetry. In the rest of the paper and for simplicity, we will consider
only the case when ${\mathcal{G}}$ is semi-simple (that is, no maximal
ideals are present in ${\mathcal{G}}$); hence the new transformation is
reducible only when $\xi$ is proportional to $F$.  In this case in the
formalism of Batalin-Vilkovosky, which suitably deals with reducible
symmetries, our symmetry is first-stage reducible.  We will apply this
formalism to quantise the extra symmetry.
\par
To obtain the dual theory, we have to perform the path integral over the
$\phi$, $A_{\mu}$ and $\Lambda$ in the action (\ref{firstaction}). There
are, therefore, two symmetries that one needs to gauge fix. The first
one is the usual local gauge transformation in (\ref{usualsym}) and the
second is the extra symmetry in (\ref{newsym}). Since the two symmetries
are completeley independent and different in nature, it is therefore
essential to keep one symmetry intact if the other is being fixed.
\par
We choose first to fix the extra symmetry in (\ref{newsym}) keeping the
gauge symmetry in (\ref{usualsym}) intact. This is easily achieved if we
choose a gauge fixing condition for the symmetry (\ref{newsym}) which
transforms covariantly with respect to the local gauge transformation
(\ref{usualsym}).
\par
We intend to employ the formalism of Batalin-Vilkovisky to quantise the
new reducible theory, we will give the main ingredients of this
formalism in what follows.

\section{Review of the Batalin-Vilkovisky Formalism}

The Batalin-Vilkovisky formalism manages theories with reducible
symmetries. The \newline Faddeev-Popov procedure is, in general, not
sufficient for such theories.  A simplistic use of the
Becchi-Rouet-Stora-Tyutin (BRST) quantisation is also inappropriate in
this case. We will give the essential tools of this formalism in that
which follows.
\par
Let ${\mathcal {S}}$ be a classical action for some generic fields
$\phi^i$, $i=1,\dots, n$ (fermionic or bosonic in nature).  The
equations of motion of this gauge action are assumed to possess at least
one solution $\phi_0$. Let $m_0$ be the number of gauge parameters
(fermionic and bosonic) of this gauge invariant action; hence $m_0$
Noether identities hold
\be
{\dd_r{\mathcal{S}}\over \dd \phi^i}R^i_{\alpha_{0}}=0
\,\,\,,\,\,\,\alpha_0=1,\dots,m_0\,\,\,.
\ee
$R^i_{\alpha_{0}}\left(\phi\right)$ are the generators of the gauge
transformations and are supposed to be regular functionals of the fields
$\phi^j$.  These transformations are written as $\delta\phi^i =
R^i_{\alpha_{0}}\delta\theta^{\alpha_{0}}$, where $\theta^{\alpha_{0}}$
are the gauge parameters.  We will denote by $\dd_r$ and $\dd_l$ the
right and left functional derivatives, respectively. We also use the de
Witt convention that summation over repeated indices includes an
integration over spacetime.
\par
The gauge symmetry is then reducible if there exists (at least on-shell)
a set of $m_1$ zero-eigenvalue eigenvectors $Z_{(1)\alpha_{1}}^
{\alpha_{0}}$ such that
\be
R^i_{\alpha_{0}}Z_{(1)\alpha_{1}}^{\alpha_{0}}|_{\phi_{0}}=0
\,\,\,\,,\,\,\,\,
\alpha_1=1,\dots,m_1\,\,\,.
\ee
The symmetry is said to be first-stage reducible if the 
null vectors $Z^{\alpha_{0}}_{(1)\alpha_{1}}$ are independent.
We will consider here only symmetries such as these. 
\par
The fields $\phi^i$ are part of a larger set of fields $\Phi^A$,
$A=1,\dots,N$ (the rest of the fields being the different
ghosts and some Lagrange multipliers necessary for gauge
fixing).
The Batalin-Vilkovisky formalism associates with each
field $\Phi^A$ an anti-field $\Phi^*_A$ possessing opposite statistics.
These anti-fields are just tools for constructing a BRST
invariant action. 
If we denote by $\epsilon\left(\Phi^A\right)\equiv 
\epsilon_A$ the statistics of the field $\Phi^A$, then the 
fermion number of the anti-field is $\epsilon\left(\Phi_A^*\right)
= \epsilon_A +1 \left({\mathrm{mod}}2\right)$.
\par
It is then guaranteed that there exists a BRST invariant 
quantum action ${\mathbf {S}}\left(\Phi,\Phi^*\right)$
which satisfies the two requirements \cite{bata85}
\bea
&{}&{\mathbf{S}}\left(\Phi,\Phi^*\right)|_{\Phi^{*}=0}
={\mathcal{S}}\left(\phi\right)\nonumber\\ &{}&\left({\mathbf
{S}},{\mathbf {S}}\right) \equiv {\dd_r {\mathbf{S}}\over
\dd\Phi^A}{\dd_l {\mathbf{S}}\over \dd\Phi^*_A} -{\dd_r
{\mathbf{S}}\over \dd\Phi^*_A} {\dd_r {\mathbf{S}}\over
\dd\Phi^A}=0\,\,\,,
\eea
The first expression demands that one can retrieve thet correct
classical field theory.  The second equation is what is known as the
master equation and its solution will be our main concern.
\par
The minimum number of fields contained within a first-stage reducible
theory is the number of fields in
$\Phi^A_{\mathrm{min}}=\left\{\phi^i,C_{(0)}^{\alpha_{0}},
C_{(1)}^{\alpha_{0}}\right\}$ plus $\Phi^*_{\mathrm{min}}$. The fields
$C_{(0)}^{\alpha_{0}}$ are assigned a ghost number equal to $1$ and are
the usual Faddeev-Popov ghosts, whilst $C_{(1)}^{\alpha_{1}}$ are the
ghosts-for-ghosts fields and have ghost number equal to $2$.  Of course,
the field $\phi^i$ has zero ghost number. The statistics of a field, or
anti-field, is the sum of the statistics of its index and the absolute
value of its ghost number.  The first stage in constructing a BRST
invariant theory is to associate an action
${\mathbf{S}}\left(\Phi_{\mathrm{min}},\Phi^*_{\mathrm{min}}
\right)$ with this minimum set of fields. This action can be expanded 
in powers of the anti-fields, where each term in the expansion has zero
ghost number. The leading terms in this expansion are of the form
\cite{bata85}
\bea
{\mathbf{S}}\left(\Phi_{\mathrm{min}},\Phi^*_{\mathrm{min}}\right) &=&
{\mathcal{S}} + \phi^*_i R^i_{\alpha_{0}} C_{(0)}^{\alpha_{0}} +
C^*_{{(0)\alpha_{0}}}\left[Z^{\alpha_{0}}_{(1)\alpha_{1}}C^{\alpha_{1}}_
{(1)} + T^{\alpha_{0}}_{\beta_{0}\gamma_{0}} C^{\gamma_{0}}_{(0)}
C^{\beta_{0}}_{(0)}\right]\nonumber\\ &+&
C^*_{(1)\alpha_{1}}\left[A^{\alpha_{1}}_{\beta_{1}\alpha_{0}}
C^{\alpha_{0}}_{(0)}C_{(1)}^{\beta_{1}} +
F^{\alpha_{1}}_{\alpha_{0}\beta_{0}\gamma_{0}}
C^{\gamma_{0}}_{(0)}C_{(0)}^{\beta_{0}}C_{(0)}^{ \alpha_{0}}\right]
\nonumber\\&+&
\phi^*_i\phi^*_j\left[B^{ji}_{\alpha_{1}} C_{(1)}^{\alpha_{1}}
+ E^{ji}_{\alpha_{0}\beta_{0}} C^{\beta_{0}}_{(0)}
C^{\alpha_{0}}_{(0)}\right]\nonumber\\
&+&
2C^*_{(0)\alpha_{0}}\phi^*_i\left[
G^{i\alpha_{0}}_{\alpha_{1}\beta_{0}} C^{\beta_{0}}_{(0)}
C^{\alpha_{1}}_{(1)} +
D^{i\alpha_{0}}_{\beta_{0}\gamma_{0}\delta_{0}}
C^{\delta_{0}}_{(0)}C_{(0)}^{\gamma_{0}}C_{(0)}^{\beta_{0}}\right]
+\dots\,\,\,,
\label{masterexpansion}
\eea
There are no more terms in this expansion for the usual first-stage
reducible theories.
\par
The master equation then imposes the following conditions on the
different coefficients in the above expansion
\bea
&{}&{\dd_r{\mathcal{S}}\over \dd\phi^i}R^i_{\alpha_{0}}C^{ \alpha_{0}}
_{(0)}=0\,\,\,\,,
\label{master1}
\\
&{}&
R^i_{\alpha_{0}} Z^{\alpha_{0}}_{(1)\beta_{1}}C_{(1)}^{\beta_{1}}
-2{\dd_r{\mathcal{S}}\over \dd\phi^j}B_{\beta_{1}}^{ji}
C_{(1)}^{\beta_{1}}\left(-1\right)^{\epsilon_{i}}=0\,\,\,\,,
\label{master2}
\\
&{}&
{\dd_r R^i_{\alpha_{0}} C^{\alpha_{0}}_{(0)}
\over\dd\phi^j} R^j_{\beta_{0}}C_{(0)}^{\beta_{0}}
+
R^i_{\alpha_{0}}T^{\alpha_{0}}_{\beta_{0}\gamma_{0}}
C_{(0)}^{\gamma_{0}}C_{(0)}^{\beta_{0}}
-
2{\dd_r{\mathcal{S}}\over \dd\phi^j}E_{\beta_{0}\gamma_{0}}^{ji}
C_{(0)}^{\gamma_{0}}C_{(0)}^{\beta_{0}}\left(-1\right)
^{\epsilon_{i}}=0\,\,\,\,,
\label{master3}
\\
&{}&
{\dd_r T^{\alpha_{0}}_{\beta_{0}\gamma_{0}} C^{\gamma_{0}}_{(0)}
C^{\beta_{0}}_{(0)}
\over\dd\phi^j} R^j_{\delta_{0}}C_{(0)}^{\delta_{0}}
+
2T^{\alpha_{0}}_{\beta_{0}\gamma_{0}}C^{\gamma_{0}}_{(0)}
T^{\beta_{0}}_{\delta_{0}\mu_{0}}C^{\mu_{0}}_{(1)}
C^{\delta_{0}}_{(0)}
+
Z^{\alpha_{0}}_{(1)\beta_{1}}F^{\beta_{1}}_{\beta_{0}\gamma_{0}
\delta_{0}}C^{\delta_{0}}_{(0)}
C_{(0)}^{\gamma_{0}}C_{(0)}^{\beta_{0}}\nonumber\\
&+&
2{\dd_r{\mathcal{S}}\over \dd\phi^j}D_{\beta_{0}\gamma_{0}\delta_{0}}
^{j\alpha_{0}}
C_{(0)}^{\delta_{0}}C_{(0)}^{\gamma_{0}}
C_{(0)}^{\beta_{0}}\left(-1\right)
^{\epsilon_{\alpha_{0}}}=0\,\,\,\,,
\label{master4}
\\
&{}&
{\dd_r Z^{\alpha_{0}}_{(1)\beta_{1}} C^{\beta_{1}}_{(1)}
\over\dd\phi^j} R^j_{\delta_{0}}C_{(0)}^{\delta_{0}}
+
2T^{\alpha_{0}}_{\beta_{0}\gamma_{0}}C^{\gamma_{0}}_{(0)}
Z^{\beta_{0}}_{(1)\delta_{1}}C^{\delta_{1}}_{(1)}
+
Z^{\alpha_{0}}_{(1)\beta_{1}}A^{\beta_{1}}_{\gamma_{1}\beta_{0}}
C^{\beta_{0}}_{(0)}
C_{(1)}^{\gamma_{1}}\nonumber\\
&+&
2{\dd_r{\mathcal{S}}\over \dd\phi^j}G_{\gamma_{1}\beta_{0}}
^{j\alpha_{0}}
C_{(0)}^{\beta_{0}}C_{(1)}^{\gamma_{1}}
\left(-1\right)
^{\epsilon_{\alpha_{0}}}=0\,\,\,\,.
\label{master5}
\eea
Here $\epsilon_i=\epsilon\left(\phi^i\right)$, whilst
$\epsilon_{\alpha_{0}}$ is the Grassmann parity of the gauge parameter.
\par
The minimum sets of fields $\Phi_{\mathrm{min}}$ and of anti-fields
$\Phi^*_{\mathrm{min}}$ can be enlarged to include more fields and their
corresponding anti-fields.  The master equation implies that, if
${\mathbf{S}}\left(\Phi_{\mathrm{min}},\Phi^*_{\mathrm{min}}\right)$ is
a solution, then
\be
{\mathbf{S}}\left(\Phi,\Phi^*\right)=
{\mathbf{S}}\left(\Phi_{\mathrm{min}},\Phi^*_{\mathrm{min}}\right)
+ {\bar {C}}_{(0)}^{*\alpha_{0}} \Pi_{(0)\alpha_{0}}
+ {\bar {C}}_{(1)}^{*\beta_{1}} \Pi_{(1)\beta_{1}}
+ C^{'*}_{(1)\beta_{1}} \Pi_{(1)}^{'\beta_{1}}
\label{fullaction}
\ee
is also a solution.  The new fields may be employed in gauge fixing as
we will see shortly, and are assigned the ghost numbers
\bea
&{}&{\mathrm{gh}}\left(\Pi_{(0)\alpha_{0}}\right)=
{\mathrm{gh}}\left(C_{(1)}^{'\alpha_{1}}\right)=0
\nonumber\\
&{}&{\mathrm{gh}}\left(C_{(0)}^{\alpha_{0}}\right)=
-{\mathrm{gh}}\left({\bar{C}}_{(0)\alpha_{0}}\right)=
-{\mathrm{gh}}\left(\Pi_{(1)\alpha_{1}}\right)=
{\mathrm{gh}}\left(\Pi_{(1)}^{'\alpha_{1}}\right)=1
\nonumber\\
&{}&
{\mathrm{gh}}\left(C_{(1)}^{\alpha_{1}}\right)=
-{\mathrm{gh}}\left({\bar{C}}_{(1)\alpha_{1}}\right)=2\,\,\,.
\eea
The fields with a star denote their corresponding anti-fields.
\par
The anti-fields are not physical fields and should be eliminated from
the theory.  This is achieved through the introduction of what is known
as the gauge-fixing fermion $\Psi\left(\Phi\right)$. This is a
functional of odd statistics and having a ghost number equal to
$-1$. The anti-fields in the full action (\ref{fullaction}) are then
replaced by
\be
\Phi^*_A ={\dd\Psi\over \dd\Phi^A}\,\,\,.
\label{gaugefixing}
\ee
\par
The functional $\Psi$ has to satisfy certain conditions in order to make
all the ghost propagators invertible.  The simplest choice of functional
$\Psi$ for first-stage reducible theories takes the form
\be
\Psi\left(\Phi\right)= 
{\bar{C}}_{(0)\alpha_{0}} \chi^{\alpha_{0}}
+ 
{\bar{C}}_{(1)\beta_{1}}\Omega^{\beta_{1}}_{\alpha_{0}}
C_{(0)}^{\alpha_{0}} 
+ 
{\bar{C}}_{(0)\alpha_{0}}\Sigma^{\alpha_{0}}_{\beta_{1}}
C_{(1)}^{'\beta_{1}}\,\,\,,
\label{gaugefermion}
\ee
where $\chi^{\alpha_{0}}\left(\phi^i\right)$ is an admissible gauge
condition for the classical fields $\phi^i$.  The matrices
$\Omega_{\alpha_{0}}^{\beta_{1}}$ and $\Sigma^{\alpha_{0}}_{\beta_{1}}$
are some suitable maximal rank matrices which remove the degeneracy of
the kinetic term of the ghosts $C_{(0)}^{\alpha_{0}}$ and
${\bar{C}}_{(0)\alpha_{0}}$.
\par
Note that the integration in the path integral over the $\Pi$'s of
(\ref{fullaction}) leads to three sets of gauge conditions. These
conditions are in the form of $\delta$-functions. To obtain the usual
quadratic gauge-fixing Lagrangian (the 't Hooft method), a linear term
in the $\Pi$'s is added to $\Psi$.  In the simplest cases the following
gauge fermion leads to to a quadratic gauge-fixing Lagrangian
\be
\widetilde{\Psi} =\Psi
+{1\over 2}\left[{\bar{C}}_{(0)\alpha_{0}}\Gamma^{\alpha_{0}
\beta_{0}}\Pi_{(0)\beta_{0}} +
{\bar{C}}_{(1)\alpha_{1}}\Theta^{\alpha_{1}}_
{\beta_{1}}\Pi_{(1)}^{'\beta_{1}} -
\left(-1\right)^{\epsilon_{\alpha_{1}}}
\Pi_{(1)\alpha_{1}}\Theta^{\alpha_{1}}_{
\beta_{1}} C_{(1)}^{'\beta_{1}}\right]\,\,\,,
\label{linearinpi}
\ee
where $\Psi$ is given in (\ref{gaugefermion}) and
$\Gamma^{\alpha_{0}\beta_{0}}$ and $\Theta^{\alpha_{1}}_{\beta_{1}}$ are
some invertible matrices assumed to contain no derivatives.  The
integration over the $\Pi$'s will give Gaussian averages of gauge
conditions instead of $\delta$-functions.  This issue will be of
considerable relevance when we consider non-Abelian duality in sigma
models.
\par
To end this brief review of the Batalin-Vilkovisky formalism, we provide
a means to determine the BRST transformations of the different fields. A
generic quantity $P\left(\Phi,\Phi^*\right)$ having statistics
$\epsilon_P$, has a BRST transformation given by
\be
\delta P=\left(-1\right)^{\epsilon_{P}}\left(P,
{\mathbf{S}}\right)\,\,\,.
\ee
This transformation is nilpotent $\left(\delta^2P=0\right)$ by virtue of
the master equation satisfied by ${\mathbf{S}}$. This definition of the
BRST transformation guarantees that ${\mathbf{S}}$ is, by construction,
BRST invariant. The factor $\left(-1\right)^{\epsilon_P}$ has been
chosen to enforce graded Leibniz rules for $\delta$.
\par
Upon elimination of the anti-fields through (\ref{gaugefixing}), the
action ${\mathbf{S}}\left(\Phi,
\Phi^*={\dd \Psi\over \dd \Phi}\right)$ is still BRST invariant.
In general, however, the nilpotency of the BRST transformation holds
only when the equations of motion of the quantum action
${\mathbf{S}}\left(\Phi,
\Phi^*={\dd \Psi\over \dd \Phi}\right)$ are used.
\par
We are now at a stage where we can apply the Batalin-Vilkovisky
formalism to theories of the form given in (\ref{firstaction}).

\section{Application of the Batalin-Vilkovisky Formalism}

In order to become familiar with the general ideas of the anti-field
formalism, let us start by quantising the action (\ref{firstaction}). We
will deal with the symmetry (\ref{newsym}) leaving the usual gauge
symmetry (\ref{usualsym})untouched throughout the procedure.  This may
be regarded as a preliminary exercise before one tackles more
complicated cases.
\par
The variation of this action with respect to $\Lambda$ leads to the
equation of motion
\be
F^a\equiv\epsilon^{\mu\nu}F^a_{\mu\nu}=0\,\,\,,
\ee
where we have written $A_\mu=A_\mu^aT_a$, $F_{\mu\nu}=F^a_{\mu\nu}T_a $
and $\Lambda=\Lambda^a T_a$. The $T_a$ are the generators of the Lie
algebra ${\mathcal{G}}$ such that
$\left[T_a\,\,,\,\,T_b\right]=f^c_{ab}T_c$.
\par
The set of classical fields is $\phi^i=\left\{\varphi,
A^a_\mu,\Lambda^a\right\}$.  The transformation we are dealing with is
Abelian and closes off-shell; hence the structure constants
$T^{\alpha_{0}}_{
\beta_{0}\gamma_{0}}$ vanish.  Let us now investigate  
which of the coefficients of the expansion (\ref{masterexpansion})
survive in this case.
\par
The tranformation (\ref{newsym}) leads to $R^i_{\alpha_{0}}$ which are
nonzero only when the index $i$ refers to the field $\Lambda^a$
\be
R^{a(x)}_{b(y)}=f^a_{bc}F^c\left(x\right)
\delta\left(x-y\right)\,\,\,,
\ee
where the index $i=\left\{a,x\right\}$ and $\alpha_{0}=\left\{
b,y\right\}$. Due to the anti-symmetry of the structure constants
$f^a_{bc}$, the null vectors of $R^i_{\alpha_{0}}$ are given by
\be
Z_{(1)(z)}^{b(y)}= F^b\left(y\right)
\delta\left(y-z\right)\,\,\,,
\ee
where the index $\beta_{1}=\left\{z\right\}$. It is clear that these
null vectors are linearly independent off-shell.; hence this theory is
said to be firts--stage reducible.  Since
$T^{\alpha_{0}}_{\beta_{0}\gamma_{0}}$, $R^i_{\alpha_{0}}$ and
$Z_{(1)\beta_{1}}^{\alpha_{0}}$ do not depend on the field $\Lambda^a$,
a solution to the master equation is obtained by setting all the other
coefficients in (\ref{masterexpansion}) to zero.

 Hence, keeping the Batalin-Vilkovisky notation, we are left with
\be
{\mathbf{S}}\left(\Phi_{\mathrm{min}},\Phi^*_{\mathrm{min}}\right)=
{\mathcal{S}}\left(\phi\right) + \phi^*_i R^i_{\alpha_{0}}
C_{(0)}^{\alpha_{0}} + C^*_{{(0)\alpha_{0}}}
Z^{\alpha_{0}}_{(1)\alpha_{1}}C^{\alpha_{1}}_ {(1)}
\,\,\,.
\ee
The full quantum action is then written in the suggestive form
\bea
{\mathbf{S}}\left(\Phi,\Phi^*\right) &=&
I\left(\phi,A,\Lambda\right)
+{\mathcal{S}}_{\mathrm{ghost}}+
{\mathcal{S}}_{\mathrm{gauge}} 
\nonumber\\
{\mathcal{S}}_{\mathrm{ghost}}&\equiv& {\dd \Psi\over \dd\Lambda^i}
R^i_{\alpha_{0}} C_{(0)}^{\alpha_{0}} + {\dd\Psi\over\dd
C_{(0)}^{\alpha_{0}}} Z^{\alpha_{0}}_{(1)\alpha_{1}}C^{\alpha_{1}}_
{(1)}
\nonumber\\
{\mathcal{S}}_{\mathrm{gauge}}&\equiv&
 {\dd\Psi\over \dd {\bar {C}}_{(0)\alpha_{0}}} \Pi_{(0)\alpha_{0}}
+{\dd\Psi\over \dd {\bar {C}}_{(1)\beta_{1}}} \Pi_{(1)\beta_{1}}
+{\dd\Psi\over \dd C_{(1)}^{'\beta_{1}}} \Pi_{(1)}^{'\beta_{1}}
\,\,\,. \label{suggestive}
\eea
The anti-fields have been eliminated using the gauge-fixing fermion
$\Psi$.
\par
The next step in determining the full quantum action is to construct the
gauge-fixing fermion $\Psi$.  As mentioned earlier, we would like to
gauge fix the transformation (\ref{newsym}) without breaking the usual
gauge symmetry in (\ref{usualsym}). This can be achieved by choosing a
gauge fixing condition which transforms covariantly under
(\ref{usualsym}).  A gauge fixing condition which has this property is
given by
\be
\chi^a = f^a_{bc}\Lambda^b F^c \,\,\,\,.
\ee
This is a set of $\left[{\mathrm{dim}}{\mathcal{G}}-
{\mathrm{rank}}{\mathcal{G}}
\right]$ equations which are compatible with the transformation 
(\ref{newsym}). The gauge fermion then takes the form
\be
\Psi =\int {\mathrm{d}}^2x \left[
{\bar{C}}_{(0) a} f^a_{bc}\Lambda^b F^c
+ 
{\bar{C}}_{(1)}\Omega_a
C_{(0)}^{a} 
+ 
{\bar{C}}_{(0) a} \Sigma^a 
C_{(1)}\right]\,\,\,.
\label{fermion1}
\ee
Under the gauge transformations (\ref{usualsym}), the ghost fields are
obviously required to transform in the adjoint representation of
${\mathcal{G}}$.  The matrices $\Omega^{\beta_{1}}_{\alpha_{0}}$ and
$\Sigma^{\alpha_{0}}_{\beta_{1}}$ are chosen such that the gauge
covariance (\ref{usualsym}) is maintained. These matrices are also
assumed to be independent of the Lagrange multiplier field, $\Lambda$.
\par
The ghost action is therefore given by
\be
{\mathcal{S}}_{\mathrm{ghost}} =
\int {\mathrm{d}}^2x\left[
{\bar{C}}_{(0)a}f^a_{bc} f^b_{de}F^cF^e
C_{(0)}^{d}
+
{\bar{C}}_{(1)}\Omega_a F^a C_{(1)}
\right]\label{ghostly}
\ee
It is clear that ${\mathcal{S}}_{\mathrm{ghost}}$ is invariant under
\bea
C_{(0)}^a &\longrightarrow& C_{(0)}^a + \alpha F^a\nonumber\\
{\bar{C}}_{(0)_a}&\longrightarrow& C_{(0)_a} + {\bar{\alpha}}
\eta_{ab} F^b\,\,\,\,
\label{ghostsym}
\eea
where $\alpha$ and $\bar{\alpha}$ are two local Grassmanian parameters.
In this sense the ghost action is degenerate (that is, the gauge fixing
did not remove all the symmetries of our theory). It is the role of the
gauge fixing Lagrangian to remove all the degeneracies.
\par
 The integration over the $\Pi$'s in $\mathcal{S}_{\mathrm{gauge}}$
leads to three conditions
\be
f^a_{bc}\Lambda^b F^c + \Sigma^a C_{(1)}^{'} =0\,\,\,\,,\,\,\,\,
\Omega_a C_{(0)}^a =0\,\,\,\,\,\,\,\,\,
{\bar{C}}_a \Sigma =0\,\,\,.
\label{3gauges}
\ee
The first condition fixes the gauge transformation in (\ref{newsym}) and
eliminates $C_{(1)}^{'}$.  Multiplication by $\eta_{ad}F^d$ of the first
equation yields $\eta_{ad}F^d\Sigma^aC_{(1)}^{'}=0$. This is sufficient
to eliminate $C_{(1)}^{'}$ provided that $\eta_{ad}F^d\Sigma^a$ does not
vanish identically. The remaining two conditions fix the ghost
transformation mentioned in (\ref{ghostsym}). We found that the two
matrices
\be
\Omega_a = \eta_{ab}F^b \,\,\,\,,\,\,\,\,\,
\Sigma^a= F^a 
\ee
satisfy all the above mentioned requirements.
\par
In this way we have constructed a BRST invariant quantum theory.  If one
wishes to eliminate the anti-fields using the gauge fermion $\Psi$ then
the BRST transformations are given by
\be
\delta_{\Psi}\Phi^{A}=(-1)^{\epsilon_{A}}\left.
\frac{\dd_{l}\mathcal{S}}{\dd\Psi^{*}_{A}}
\right|_{\Phi^{*}=\frac{\dd\Psi}{\dd\Phi}}
\ee
It is then a simple matter to write down the BRST transformations for
the fields
\bea
\delta_{\Psi}\Lambda^{a}&=&f^{a}_{bc}F^{c}C^{b}_{(0)} \nonumber\\
\delta_{\Psi}C^{a}_{(0)}&=&-F^{a}C_{(1)}\nonumber\\
\delta_{\Psi}\bar{C}_{(0)\,a}&=&-\Pi_{(0)\,a}\nonumber\\
\delta_{\Psi}\bar{C}_{(1)}&=&\Pi_{(1)}\nonumber\\
\delta_{\Psi}C^{'}_{(1)}&=& 
\delta_{\Psi}\Pi_{(0)\,a}=\delta_{\Psi}\Pi_{(1)}
= \delta_{\Psi}\Pi^{'}_{(1)}=0\; .
\eea
It then follows that the BRST transformations are nilpotent.
\par
Finally, we would like to investigate a point which is relevant to
non-Abelian duality. This concerns the addition of linear terms in the
$\Pi$'s to the gauge fermion $\Psi$.  In this case the new gauge fermion
takes the form
\be
\widetilde{\Psi} =\Psi
+{1\over 2}\int{\mathrm{d}}^2x
\left[{\bar{C}}_{(0)a}\Gamma^{ab}
\Pi_{(0) b} +
{\bar{C}}_{(1)}^{'}\Theta \Pi_{(1)}^{'} -
\Pi_{(1)}\Theta C_{(1)}\right]\,\,\,,
\ee
where $\Psi$ is the gauge fermion given in (\ref{fermion1}). In order to
maintain covariance under (\ref{usualsym}), a simple choice for the two
matrices $\Gamma^{\alpha_{0}\beta_{0}}$ and
$\Theta^{\alpha_{1}}_{\beta_{1}}$ is
\be
\Gamma^{ab}= n\eta^{ab}\,\,\,\,\,,\,\,\,\,\,
\Theta = m\,\,\,,
\ee
where $\eta^{ab}$ is the inverse of $\eta_{ab}$ and $n$ and $m$ are two
constant parameters.
\par
The integration over the $\Pi$'s results in the quadratic gauge-breaking
Lagrangian
\bea
{\mathcal{S}}_{\mathrm{gauge}} &=&
\int {\mathrm{d}}^2x \left[
-{1\over 2n}\left(f^a_{bc}\Lambda^bF^c\right)\eta_{ad}
             \left(f^d_{rs}\Lambda^rF^s\right)
-{1\over m} {\bar{C}}_{(0)a}F^a\eta_{bc}F^c C^b_{(0)} 
\right.\nonumber\\
&-&\left.{1\over 2n}C^{'}_{(1)}F^a\eta_{ab}F^bC^{'}_{(1)}\right]
\,\,\,.
\eea
This is the usual Gaussian gauge fixing Lagrangian. The first term
removes the gauge freedom of the original action while the second term
removes the degeneracy of the ghost Lagrangian (\ref{ghostly}).  The
last term is required for BRST invariance and is a characteristic of the
anti--field formalism.
\par
This completes the quantisation of the new symmetry (\ref{newsym}).  Let
us now list the consequences of our work on non--Abelian duality.  We
will, however, leave the detailed investigation to a forthcoming
publication \cite{us}.

\section{Conclusions}

We have shown in this paper that the procedure by which non--Abelian
duality is implemented in sigma models naturally leads to the presence
of a reducible symmetry.  We have dealt with this symmetry using the
Batalin--Vilkovisky formalism.  This unavoidably introduces new fields
into the theory.  Some of these fields are bosonic in nature
$\left(\,C_{(1)},\,\bar{C}_{(1)}\, \mbox{and}\,C^{'}_{(1)}\right)$ and
could play a r\^{o}le similar to that of the Lagrange multiplier
$\Lambda$.  This is further investigated in \cite{us}.
\par
In order to proceed further in the determination of the dual theory one
must carry out an integration over the gauge fields in the full action
(\ref{suggestive}).  However, this is no more straightforward as this
action includes terms quadratic in the field strength of the gauge
fields.  This fact is worsened if we consider the gauge fermion
$\tilde{\Psi}$ instead of $\Psi$.  The integration over the gauge fields
would lead to a dual theory containing non--local terms.  The latter can
no longer be interpreted as a sigma model corresponding to a string
background.  This issue, in fact, is particularly specific to our choice
of gauge fixing condition which contains the field strength.  It is
possible to find a gauge breaking term which does not contain any gauge
fields.  These types of gauges are reported in \cite{us} and involve
only the sigma model fields $\varphi$ and the Lagrange multiplier
$\Lambda$.
\par
In this paper we have started by quantising the symmetry (\ref{newsym})
keeping manifest the usual gauge symmetry (\ref{usualsym}).  It is then
natural to address the following question: could we have started the
other way around?  That is, to quantise first the symmetry in
(\ref{usualsym}) . This is an important issue which is also investigated
in the forthcoming paper \cite{us}.  Let us simply mention that there
are two ways in which to gauge fix the symmetry (\ref{usualsym}).  The
first is, for instance, to choose a standard gauge of the Landau type
$\partial^{\mu}A^{a}_{\mu}=0$.  This could be solved by setting
$A^{a}_{\mu}=\epsilon_{\mu\nu}\partial^{\nu}\lambda^{a}$ and leads to a
non--vanishing field strength.  Therefore, this type of gauge fixing
does not break the new symmetry in (\ref{newsym}).  The second type of
gauge fixing is a non--standard one and involves setting some fields
($\varphi\,\mbox{and}\,\Lambda$) to zero.  In general, however, this
gauge automatically breaks the new symmetry in (\ref{newsym}).  This is
the type of gauge fixing which has been considered in the literature on
non--Abelian duality.
\par
Another direction of research concerns a class of sigma models,
identified in \cite{nouri96}, which possess a symmetry of the type
considered in this paper.  It would be interesting to explore their
their quantisation \`{a} la Batalin--Vilkovisky.  This is a quite
involved programme as the Noether currents $R^{i}_{\alpha_{0}}$ and the
null vectors $Z^{\alpha_{0}}_{(1)\,\alpha_{1}}$ are field dependent, and
the gauge algebra closes only on--shell.  We will report on the work in
progress in \cite{us}.

\subsection*{Acknowledgements}

We would like to thank Tim Morris, Richard Moss, Douglas Ross and
Caroline Wilkins for discussions.

\end{document}